\documentclass[prc,showpacs,showkeys,superscriptaddress,nofootinbib,twocolumn,floatfix,noeprint,nolongbibliography]{revtex4-1}
\usepackage{amsfonts}
\usepackage{float}
\usepackage{placeins}
\usepackage{graphicx}
\usepackage[hidelinks,pdftex]{hyperref}
\usepackage{color,amsmath,amssymb,bm}

\def\blfootnote{\xdef\@thefnmark{}\@footnotetext}

\def\eg{{\em e.g.}}
\def\ie{{\em i.e.}}

\newcommand{\beq}{\begin{equation}}
\newcommand{\eeq}{\end{equation}}
\newcommand{\bea}{\begin{eqnarray}}
\newcommand{\eea}{\end{eqnarray}}

\def \vp {\bm{p}}

\begin{document}

\title{Heavy-Light Susceptibilities in a Strongly Coupled Quark-Gluon Plasma}
\author{Shuai Y.\,F. Liu}
\affiliation{Quark Matter Research Center, Institute of Modern Physics, Chinese Academy of Sciences, Lanzhou 730000, China}
\affiliation{Cyclotron Institute and Department of
Physics and Astronomy, Texas A\&M University, College Station, TX 77843-3366, USA}
\author{Ralf Rapp}
\affiliation{Cyclotron Institute and Department of
Physics and Astronomy, Texas A\&M University, College Station, TX 77843-3366, USA}

\date{\today}

\begin{abstract}
Quark number susceptibilities as computed in lattice QCD are commonly believed to provide insights into the microscopic 
structure of QCD matter, in particular its degrees of freedom.  
We generalize a previously constructed partonic $T$-matrix approach to finite chemical potential to calculate various susceptibilities, in particular 
for configurations containing a heavy charm quark. At vanishing chemical potential and moderate temperatures, this approach  
predicts large collisional widths of partons generated by dynamically formed hadronic resonance states which lead to transport parameters 
characteristic for a strongly coupled system. The quark chemical potential dependence is implemented into the propagators and the in-medium 
color potential, where two newly introduced parameters for the thermal and screening masses are fixed through a fit to the baryon number 
susceptibility, $\chi^B_2$. With this setup, 
we calculate heavy-light susceptibilities without further tuning; the results qualitatively agree with the lattice-QCD (lQCD) data for both $\chi^{uc}_{11}$ and 
$\chi^{uc}_{22}$. This implies that the lQCD results are compatible with a significant content of broad $D$-meson and charm-light diquark bound states 
in a moderately hot QGP. 
\end{abstract}
\keywords{Quark Gluon Plasma, Heavy Quarks, Ultrarelativistic Heavy-Ion Collisions}

\maketitle
\section{Introduction}
\label{sec_intro}
In ultra-relativistic heavy-ion collisions, a new state of matter -- quark-gluon plasma (QGP) -- can be created, at  temperatures of more 
than $ 10^{8} $ times the surface temperature of the sun, the hottest matter created in the laboratory to date.
The QGP is a fundamental realization of a many-body system governed by the strong nuclear force described by quantum chromodynamics (QCD). 
Its properties, as deduced from heavy-ion collision experiments to date~\cite{Shuryak:2014zxa,Braun-Munzinger:2015hba,Busza:2018rrf,Dong:2019byy}, 
suggest it to be a strongly-coupled liquid with transport properties near conjectured lower bounds set by quantum mechanics.  
From the theoretical side, first-principle information can be obtained from numerical simulations of the space-time discretized partition function of QCD at finite 
temperature, referred to as lattice QCD. At vanishing baryon chemical potential, $\mu_B$=0, lattice-QCD (lQCD) computations have achieved accurate
results for the equation of state (EoS) of QCD matter~\cite{Borsanyi:2010cj,Bazavov:2014pvz} and revealed that the transition between hadronic matter 
and the QGP is a smooth crossover~\cite{Aoki:2006we}. 
However, there are several quantities that are currently not accessible to lQCD computations and are not straightforward to extract from 
euclidean space-time. Nevertheless, 
high-quality lQCD ``data" provide valuable benchmarks and insights for microscopic model calculations that, in turn, can be deployed to the phenomenology of 
heavy-ion collisions. 

The strategy of utilizing lQCD data as ``numerical experimental data" for model building has been widely applied in the literature, including 
calculations of the EoS~\cite{Levai:1997yx,Peshier:2005pp,Plumari:2011mk}, quarkonium correlation 
functions~\cite{Wong:2004zr,Mocsy:2005qw,Cabrera:2006wh,Alberico:2006vw,Mocsy:2007yj,Riek:2010py} and/or heavy-quark (HQ) free 
energies~\cite{Liu:2015ypa,Rothkopf:2019ipj}. Taking advantage of the progress in 
lQCD~\cite{Aarts:2007pk,Aarts:2011sm,Petreczky:2004pz,Kaczmarek:2005ui,Borsanyi:2010cj,Bazavov:2014pvz}, 
we have developed a thermodynamic $T$-matrix approach~\cite{Liu:2016ysz,Liu:2017qah} which was rooted in three sets of lQCD data: the HQ free 
energy, Euclidean quarkonium correlator ratios, and the EoS for $N_f$=2+1 light-quark flavors. 
The lQCD results were instrumental in constraining the input parameters for the $T$-matrix approach, in particular its in-medium driving kernel and 
the effective thermal-parton masses,  and subsequently enabled controlled studies of spectral and transport properties of the 
QGP~\cite{Liu:2016ysz,Liu:2018syc,Liu:2020dlt}. 

In addition, quark-number susceptibilities~\cite{Borsanyi:2011sw,Bazavov:2012jq,Bellwied:2015lba,Ding:2015fca} -- derivatives of the partition 
function with respect to chemical potentials of different quantum numbers such as baryon number, isospin and/or strangeness -- have proven to be a 
rich source of information for effective models (see, \eg, Ref.~\cite{Ratti:2018ksb} for a recent review). They can probe aspects of the chiral transition, 
the EoS (\eg, its hadron-chemistry and extension to finite $\mu_B$) and its fluctuation properties, related to the effective degrees of freedom of the charge 
carriers. In the present paper, we are mostly interested in the latter aspect in the context of the thermodynamic $T$-matrix approach mentioned above. 
For moderate QGP temperatures, it predicts the emergence of broad hadronic bound states whose role in the EoS gradually increases as the pseudo-critical 
temperature ($T_{\rm pc}$) is approached from above. These states, which are generated from a ladder resummation of the in-medium interaction kernel, 
play a key role in producing large (resonant) interaction strength for elastic parton scattering which entail transport parameters characteristic for a strongly 
coupled system~\cite{Liu:2016ysz}. The diagonal charm-quark susceptibility, $\chi^c_2$, has been calculated previously in the $T$-matrix approach in 
Ref.~\cite{Riek:2010py}, where the free and internal energies were used as potential proxies. In particular, it has been found that sizable charm-quark
widths, $\Gamma_c\simeq100-200$\,MeV, can lead to a significant enhancement over the zero-width quasiparticle result. 

Besides the diagonal susceptibilities,  $ x^c_{2,4}$, lQCD computations are also available for off-diagonal heavy-light combinations, $\chi^{uc}_{11}$ and 
$\chi^{uc}_{22}$~\cite{Bazavov:2014yba,Mukherjee:2015mxc}. Model calculations have thus far focused on off-diagonal susceptibilities in the 
$N_f$=2+1 sector, \ie, $ \chi^{us}_{11} $ or $ \chi^{ud}_{11} $. In perturbative hard-thermal loop (HTL) calculations~\cite{Haque:2014rua}, the latter 
have been found to vanish, but they are expected to become non-vanishing at order $g^6$ (including an additional logarithmic 
dependence)~\cite{Blaizot:2001vr}.  Within Polyakov loop-extended Nambu–Jona-Lasinio (PNJL) models~\cite{Ratti:2011au,Bhattacharyya:2016jsn} 
and  the hadron resonance gas (HRG) model~\cite{Karthein:2021cmb,Miyahara:2017eam},  $ x^{us}_{11} $ has been found to be negative, in 
agreement with lQCD data, indicating the importance of hadronic degrees of freedom in the vicinity of $T_{\rm  pc}$. The analysis of the off-diagonal 
heavy-light susceptibilities from lQCD~\cite{Bazavov:2014yba,Mukherjee:2015mxc} using a schematic model of a mixture of HRG and 
free charm-quark degrees of 
freedom suggest a similar interpretation in the charm sector. The formation of heavy-light resonance states in the QGP has been put forward in earlier 
works~\cite{vanHees:2004gq,vanHees:2007me,Riek:2010fk,Liu:2018syc} as a key ingredient to evaluate the HQ diffusion coefficient, which requires
a large non-perturbative contribution in order to describe open heavy-flavor observables in heavy-ion collisions~\cite{Rapp:2018qla,He:2019vgs}. It is 
therefore important to investigate the manifestation of heavy-light correlations in the off-diagonal heavy-light susceptibilities for which very few calculations 
are available to date~\cite{Miyahara:2017eam}. In particular, we are not aware of strongly-coupled approach beyond mean-field approximations. In the
present study we employ the $T$-matrix approach, which realizes a strong-coupling scenario through the dynamical formation of hadronic states as the 
QGP temperature decreases toward  $T_{\rm pc}$.

The remainder of the ms. is organized as follows. In Sec.~\ref{sec_model}, we briefly recall the basic components of the $T$-matrix formalism as developed 
earlier and then introduce the procedure to extend it to finite $\mu_q$ and $ \mu_c $. In Sec.~\ref{sec_result}, we calculate and discuss the results of susceptibilities, 
using the light-light sector to constrain the $\mu_B$-dependent potential parameters, and then focus on the heavy-light susceptibilities and their interpretation
in the context of lQCD data. In Sec.~\ref{sec_con}, we conclude and indicate future lines of investigation.
 
\section{Thermodynamic $T$-Matrix Formalism at Finite Chemical Potential}
\label{sec_model}
The theoretical framework used in this work is thermodynamic $T$-matrix developed in Refs.~\cite{Liu:2016ysz,Liu:2017qah}. 
It is based on a Dyson-Schwinger type 
set-up for in-medium 1- and 2-body propagators, where the scattering kernel is approximated through a 3D reduction of the 4D Bethe-Salpeter equation.
This enables closed-form solutions and facilitates constraints of the in-medium potential through lQCD data for the HQ free energy. 
For the ``strongly coupled scenario"
(SCS), which we will focus on here, the in-medium potential is significantly larger than the free energy, with long-range remnants
 of the confining force surviving well above $T_{\rm pc}$ where they play a central role for the long-wavelength properties of the QGP (such as  transport 
coefficients).  A noteworthy achievement in the many-body part of that work is the full off-shell evaluation of the ladder resummation in the
Luttinger-Ward functional, which encodes the interaction contribution to the pressure of the system. The selfconsistent calculations were all carried 
out at vanishing quark chemical potential, $\mu_q$=$\mu_B$/3=0.

To calculate susceptibilities, we need to extend the $T$-matrix formalism to finite chemical potentials. In this work, we focus on the light-quark 
($\mu_q $) and charm-quark $\mu_{c}$ chemical potentials. The pertinent dependences need to be added to the propagators and the 
two-body potential, $V$, which are the two most important components of the $T$-matrix approach.
At a finite $\mu_q$ and $\mu_c$, the ``bare" quark propagators take the form
\begin{align}
&G^{0}_i(z,\vp)=\frac{1}{z-\varepsilon_{\vp} \pm\mu_i},\,\varepsilon_{\vp}=\sqrt{M_i^2+p^2}
\label{eq_prop}
\end{align}
with $i=q,c$ for light or charm quarks, respectively. For simplicity, the strange quark is treated as a light flavor degenerate with $u$ and $d$ quarks. 
The gluon propagator does not have an explicit $ \mu_i$ dependence. However, for all effective thermal-parton masses, $M_i$, we allow for an 
additional $\mu_q$ dependence as
\begin{align}
&M_i=M^0_i\sqrt{1+b_m \left(\frac{\mu_q}{T}\right)^2}
\end{align}
where the $M^0_i$ denote the temperature dependent masses at zero chemical potential which are fixed independently in Ref.~\cite{Liu:2017qah} 
by fitting the QGP EoS at $\mu_q$=0. 
The functional form of the $\mu_q$-dependence of the masses is motivated by HTL calculations at finite chemical potential~\cite{Blaizot:1999ap}; the
parameter $b_m$ will be fixed by the baryon number susceptibility, $\chi^B_2$, computed in lQCD~\cite{Borsanyi:2011sw,Bazavov:2012jq}.

For the two-body potential, $V$, our starting point is the strongly coupled solution (SCS) of Ref.~\cite{Liu:2017qah}. The only addition here is a 
$\mu_q$ dependence to the original Debye screening mass by using the ansatz
\begin{align}
m_d=m_d^0\sqrt{1+b_s \left(\frac{\mu_q}{T}\right)^2} \ , 
\end{align}
where $m_d^0$ is the $T$-dependent zero-chemical potential value fixed in Ref.~\cite{Liu:2017qah} essentially in fits to the HQ free energy. Also this ansatz 
is motivated by the HTL results of Ref.~\cite{Blaizot:1999ap} with $b_s$ being our second parameter in fitting to $\chi^B_2 $~\cite{Bazavov:2012jq}. Note that 
the screening of the string term is not independent but will be determined by $m_d$ according to the relation $m_s\propto (\sigma m_d^2)^{1/4}$, 
cf.~Sec.~D.1 of Ref.~\cite{Liu:2017qah}.

With the additional ingredients specified in the three equations above, the $T$-matrix approach is generalized to finite chemical potential. We first 
selfconsistently evaluate the coupled system of Dyson-Schwinger equations for the single-parton propagators and their 2-body $T$-matrices in all
available color channels and up to $L$=5 partial waves. Then, using the same procedure as in our original work~\cite{Liu:2017qah}, the pressure, $P$, 
can be calculated as a function of  $P(\mu_q,\mu_c)$ at finite $\mu_q$ and $\mu_c $ using the Matrix-Log resummation technique
for the Luttinger-Ward functional introduced in Refs.~\cite{Liu:2016ysz,Liu:2017qah}.  We define the dimensionless pressure as 
$\hat{P}(\hat{\mu}_q,\hat{\mu}_c)=P(\hat{\mu}_q T,\hat{\mu}_c T)/T^4$ where the $ \hat{\mu}_i=\mu_i/T $ are also dimensionless. 
The susceptibilities are obtained from the numerical derivatives of the pressure $ \hat{P} $ with respect to $\hat{\mu}_q$ and $\hat{\mu}_c$. Since 
$\mu_q=(1/3)\mu_B$, the second-order baryon number susceptibility follows from quark-number susceptibility as
\begin{align}
&\chi^{B}_2=\frac{\partial^{2}\hat{P}}{\partial\hat{\mu}_B^2}=\frac{1}{9}\frac{\partial^{2}\hat{P}}{\partial\hat{\mu}_q^2} \ . 
\end{align}
Likewise, we obtain the off-diagonal heavy-light susceptibilities as
\begin{align}
&\chi^{qc}_{nm}=\frac{\partial^{n+m}\hat{P}}{\partial\hat{\mu}_q^n\partial \hat{\mu}_c^m} \ .
\end{align}
In Ref.~\cite{Mukherjee:2015mxc},  $ \chi^{uc}_{nm} $ is explicitly shown (rather than $\chi^{qc}_{nm}$). Since the $u$, $d$ and $s$ quarks are treated 
as degenerate in our work, we have for the case $n=1$ the relation  
\begin{align}
&\chi^{uc}_{1m}=\frac{\partial^{1+m}\hat{P}}{\partial\hat{\mu}_u\partial \hat{\mu}_c^m}=\frac{1}{3}\frac{\partial^{1+m}\hat{P}}{\partial\hat{\mu}_q\partial \hat{\mu}_c^m} \  .
\end{align}
For $n=2$, we have the approximate relation
\begin{align}
&\chi^{uc}_{2m}=\frac{\partial^{2+m}\hat{P}}{\partial\hat{\mu}_u\partial \hat{\mu}_c^m}\approx\frac{1}{3}
\frac{\partial^{2+m}\hat{P}}{\partial\hat{\mu}_q^2\partial \hat{\mu}_c^m} \ , 
\end{align}
where we neglect small terms like $\chi^{udc}_{11m}$, $\chi^{usc}_{11m}$, $\chi^{usc}_{11m}$. Using $ \chi^{BC}_{mn}$ data from 
Ref.~\cite{Mukherjee:2015mxc} we have verified that these terms lead to less than 20\% difference at $T$=0.194\,GeV (the lowest $T$ in our work), 
and that they are negligible at higher temperature.
 

\section{Numerical results}
\label{sec_result}
\begin{figure} [!tb]
	\centering
	\includegraphics[width=0.95\columnwidth]{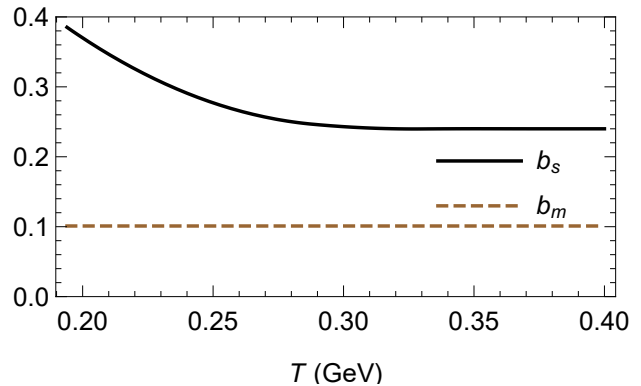}
	\caption{Results for the temperature-dependence of the parameters $b_s $and $b_m$ (introducing a $\mu_q$-dependence for the screening 
mass in the potential and for the thermal parton masses in the EoS, respectively) following from the fit shown in the upper panel of Fig.~\ref{fig_sus}.}
	\label{fig_fit}
\end{figure}

\begin{figure} [!tb]
	\centering
	\includegraphics[width=0.95\columnwidth]{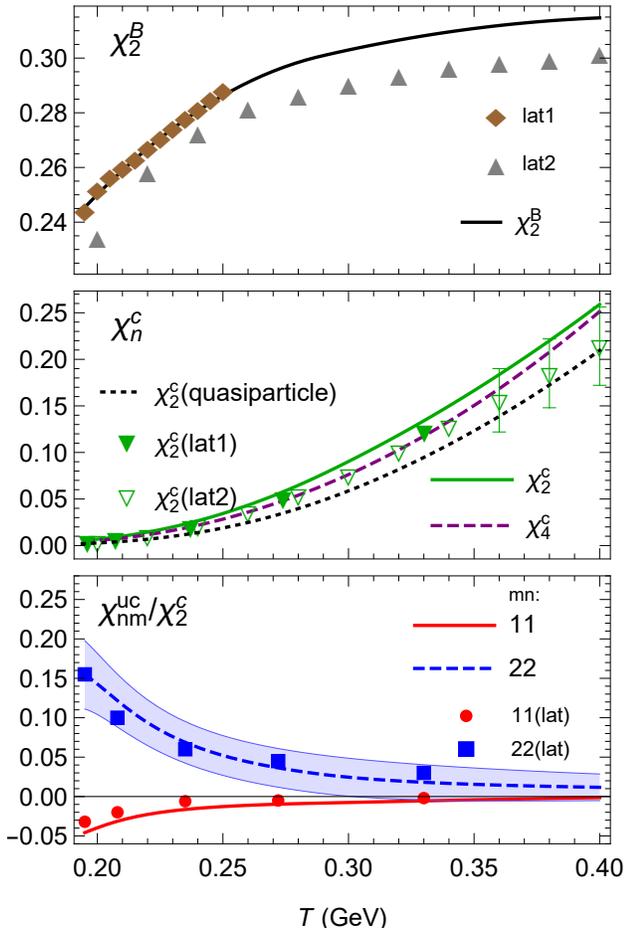}
	\caption{Upper panel: our fit to the baryon susceptibility, $ \chi^B_2 $, of the lQCD data by the Hot-QCD collaboration ("lat1")~\cite{Bazavov:2012jq}; 
also shown are the results of the Wuppertal-Budapest group ("lat2")~\cite{Borsanyi:2011sw}. 
Middle panel: our predictions for the diagonal charm susceptibilities $ \chi^c_2 $ (solid line) and $\chi_4^c$ (dashed line), compared to $N_f$=2+1 Hot-QCD and $N_f$=2+1+1 Wuppertal-Budapest lQCD data; also shown is our result when neglecting charm-quark widths and heavy-light correlations (dotted line).  Lower panel: heavy-light  
susceptibilities $\chi^{uc}_{mn}/\chi^c_2$  with ``1-$\sigma$" error band compared to Hot-QCD lQCD data~\cite{Mukherjee:2015mxc}.}
	\label{fig_sus}
\end{figure}

In our numerical evaluation of the chemical-potential derivatives, we start by selfconsistently evaluating the pressure, $\hat{P}(\hat\mu_q,\hat\mu_c)$, 
on a grid of 12 pairs of values for the light- and charm-quark chemical potentials (all combinations of $\hat\mu_q$=0,0.2 and $\hat\mu_c$=0.2,0.4,0.6,0.8), 
for a given input of the $b_m$ and $b_s$ parameters (we work in a ``quenched-charm" approximation as was done in the lQCD computations of 
Ref.~\cite{Mukherjee:2015mxc}, where charm quarks are not part of the bulk medium). 
Utilizing a polynomial ansatz for the pressure, 
\begin{align}
\label{eq_pfit}
&\hat{P}(\mu_q,\mu_c)=\hat{P}_0+\frac{\chi^{q}_2}{2}\hat{\mu}_q^2+\frac{\chi^{c}_2}{2}\hat{\mu}_c^2+\frac{\chi^{c}_4}{4!}\hat{\mu}_c^4\nonumber\\
&+\frac{\chi^{uc}_{11}}{1!\,1!}\hat{\mu}_q\hat{\mu}_c+\frac{\chi^{uc}_{22}}{2!\,2!}\hat{\mu}_q^2\hat{\mu}_c^2+\frac{\chi^{uc}_{13}}{1!\,3!}\hat{\mu}_q^2\hat{\mu}_c^2
\end{align}
(where $\hat{P}_0$ denotes the scaled pressure at vanishing chemical potnetials), we fit its coefficients to the numerically computed results from the 
$T$-matrix approach. Note that the susceptibilities defined by the derivatives to the pressure are also the coefficients of its Taylor expansion
in terms of $\mu_q$ and $\mu_c $. The fits are truncated at fourth order in $\hat{\mu}_c$ and second order in $\hat\mu_q$ 
where the latter is chosen sufficiently small to render higher orders negligible. 


We first  tune the parameters $b_m$ and $b_s$ to reproduce the baryon-number susceptibility, $\chi^B_2(T)$, computed on the lattice (where 
$\hat\mu_c$ is set to zero). Note that in our definition $\chi^B_2$ is dimensionless  (corresponding to $\chi^B_2/T^2$ in the convention where 
pressure and chemical potentials are not scaled by powers of temperature). The temperature dependence inferred for the $b$-parameters is shown 
in Fig.~\ref{fig_fit}, and the resulting baryon susceptibility in the upper panel of Fig.~\ref{fig_sus}. While for $b_m$  a constant value turns out 
to be sufficient for our purposes here, the $b_s$ parameter requires a moderate rise when approaching $T_{\rm pc}$ from above. 
We have geared our fit of $\chi^B_2$ toward the lQCD data of the Hot-QCD 
collaboration~\cite{Bazavov:2012jq} (denoted as ``lat1"), since their results have been the basis for our selfconsistent fits of the EoS in our previous 
work~\cite{Liu:2017qah}. The lQCD data for $\chi_B^2$ from the Wuppertal-Budapest group~\cite{Borsanyi:2011sw} (denoted by ``lat2")
are somewhat smaller.

\begin{figure} [!tb]
	\centering
	\includegraphics[width=0.95\columnwidth]{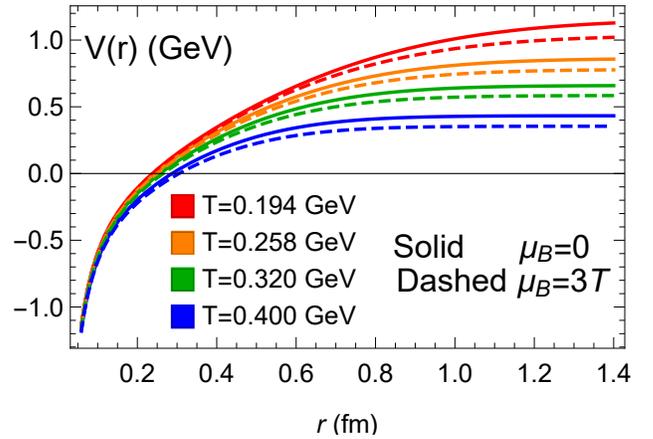}
\caption{The effects of the finite-$\mu_q$ screening on the in-medium two-body potential in the color-singlet channel resulting from our fit to the baryon 
susceptibility, $\chi_2^B$ (solid lines: $\mu_q$=0, dashed lines: $\mu_q$=$T$).}
	\label{fig_vub}
\end{figure}

The resulting $\mu_q$-dependence of the two-body potential is displayed in Fig.~\ref{fig_vub}. Since the density of the partons in the medium increases 
with $\mu_q$, the potential exhibits an expected increase in screening at fixed temperature. However, this effect appears to be relatively moderate, 
\eg, at $\mu_q/T=1$, the increase in $m_d$  amounts to only  $\sim$10-20\,\%. Recalling the relation between the color-Coulomb Debye mass and the 
screening mass of the string term, $ m_s\propto (\sigma m_d^2)^{1/4}$, and the infinite-distance value of the potential,  
$ V(r=\infty) =-(4/3)\alpha_s m_d+\sigma/m_s$~\cite{Liu:2017qah}, we find that the long-range part of the potential is only suppressed by  $\sim$5-10\,\%. 
While the predominant impact of the finite chemical potential originates from the parton propagators, Eq.~(\ref{eq_prop}), the additional 
$\mu_ q$ dependence of the potential and quark masses is essential to achieve a good fit to $\chi^B_2$. This indicates that the quark number susceptibilities 
are sensitive to microscopic physics at finite $ \mu_B$.

Next, we turn to the diagonal charm susceptibilities, $\chi_2^c$ and $\chi_4^c$,  shown in the middle panel of Fig.~\ref{fig_sus}. 
Since they are essentially independent of $\mu_q$, they are genuine predictions  of the $T$-matrix calculations at $\mu_q$=0 where all parameters 
have been fixed in our previous work~\cite{Liu:2017qah}. The result for  $\chi_2^c$ shows good 
agreement with the Hot-QCD lattice results, while the Wuppertal-Bielefeld results are somewhat lower (we recall that the latter have been computed in 
$N_f=2+1+1$-flavor QCD, \ie, including dynamical charm quarks, while our results are closer to the quenched-charm approximation as adopted in the 
Hot-QCD computations).
In particular, the $T$-matrix results are significantly larger than calculations using a quasiparticle approximation with the same charm-quark pole 
mass as in the $T$-matrix; 
in other words, the finite-width effects in the charm-quark spectral functions (with $\Gamma_c\simeq0.5$\,GeV) and the interaction effects in the 
Luttinger-Ward Functional of the pressure (including broad ``$D$"-meson and heavy-light diquark bound states for temperatures below 
$T\simeq$~250\,MeV) are essential for the 
agreement with the lQCD data. In view of this, it appears rather non-trivial that also the fourth-order derivative is in approximate agreement with the lQCD 
data which exhibit a close agreement between $\chi_2^c$ and $\chi_4^c$ (we have also verified that residual numerical uncertainties are rather 
significant in our extraction of $\chi_4^c$: \eg, when reducing the numerical tolerance from 5 to 4 digit accuracy, $\chi^4_c$ is reduced by 
ca.~10\% toward higher temperatures). 

\begin{figure} [!tb]
	\centering
	\includegraphics[width=0.95\columnwidth]{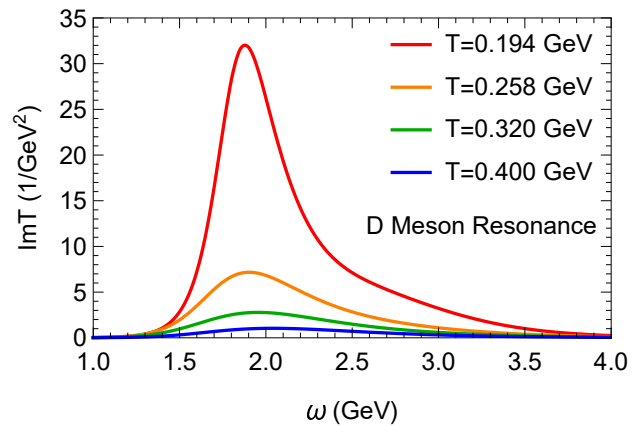}
	\caption{The imaginary part of the charm-light quark $T$-matrix in the $S$-wave color-singlet channel at four temperatures. On recognizes the 
emergence of  $D$-meson-like resonance as the temperature approaches $T_{\rm pc}$ from above.}
	\label{fig_imt}
\end{figure}

Finally, we turn to the off-diagonal susceptibilities, which are commonly normalized by $\chi_2^c$ to achieve a (partial) cancellation of the HQ 
mass effects. It turns out that the fit of the $\chi$ coefficients to the pressure, Eq.~(\ref{eq_pfit}), which we have numerically 
computed on a finite number of mesh points in the $\hat\mu_q$-$\hat\mu_c$ plane, allows for several minima where the
 deviations between fitted and calculated data are of order $10^{-5}$ or below. Since this is small compared to
our current numerical accuracy, the different minima are a priori equally likely to represent the ``true" solution.
To lift this degeneracy, we therefore impose a constraint, $ x^{uc}_{13}= x^{uc}_{11} $, motivated by lQCD data, to find the minimum
compatible with this condition (in principle, we could then release it again and find a local minimum with $ x^{uc}_{13}\approx x^{uc}_{11} $, 
but for simplicity we focus on the results with the constraint $x^{uc}_{13}= x^{uc}_{11}$). 
We reiterate that we have not  ``refit" the two ``b" parameters which were solely fixed through the light-quark susceptibilities (cf. the 2.~paragraph in
Sec~.\ref{sec_model}). Therefore, the resulting heavy-light susceptibilities can also be regarded as predictions of the model, with the numerical 
caveat outlined above. 
The pertinent results are shown in the lower panel of the Fig.~\ref{fig_sus} in terms of  $ \chi^{uc}_{11}/\chi_2^c$ and $ \chi^{uc}_{22}/\chi_2^c$,
where the error band illustrates  the 1-$ \sigma$ band of the fit, indicating that the extraction of $x^{uc}_{22}$ (and other 4th-order coefficients) is 
rather challenging in this calculation (the uncertainty is much smaller for $ \chi^{uc}_{11}$).  Again, we find a fair semi-quantitative agreement with 
lQCD data, which generally supports the role of non-perturbative physics in the QGP near $T_{\rm pc}$. The strongly-coupled features of the 
system, such as large scattering rates of the partons (which suppress the single-parton contributions) and the related onset of heavy-light 
bound-state formation (which enhance the correlated parton contributions, cf.~Fig.~\ref{fig_imt}) 
do not lead to apparent discrepancies with charm-quark susceptibilities computed in lQCD. The only other calculation we are aware of is a mean-field 
hadron-quark cross over model which predicts a positive $\chi^{uc}_{11}$~\cite{Miyahara:2017eam}, while the $\chi_{11}^{us}$ calculated in that 
work is negative and in agreement with lQCD data, thus finding no obvious connection between those two quantities. On the other hand,  the HTL 
perturbative analysis of  Ref.~\cite{Haque:2014rua} finds vanishing off-diagonal $us$ and $ud$ susceptibilities while in the PNJL calculations 
of  Refs.~\cite{Ratti:2011au,Bhattacharyya:2016jsn} the results for $x^{us}_{11}$ are negative but tend to under predict the lQCD data; in particular, 
fluctuations beyond the mean-field level were found to be essential to improve the agreement with lQCD data~\cite{Ratti:2011au}.
In a very recent HRG analysis~\cite{Karthein:2021cmb}, the inclusion of an extended set of strange-baryon resonances as predicted 
by the quark model, in combination with excluded-volume corrections, can reproduce the lQCD results up to $T\simeq$170\,MeV, where the interplay of 
mesonic and baryonic contribution, which have opposite signs~\cite{Ratti:2011au}, is critical. Our calculations also include such effects through the 
dynamical formation of mesonic and diquark resonances in the attractive color channels (color-singlet and anti-triplet, respectively; cf. Fig.~\ref{fig_imt}), 
However, the resonance correlations dissolve as temperature increases which is essential for the agreement with lQCD data at higher temperature.

We have also attempted to calculate the heavy-light susceptibilities using the ``weakly coupled solution" (WCS) of Ref.~\cite{Liu:2017qah} (where
the inter-quark potential is close ot the HQ free energy); however, the results were quantitatively rather inconclusive (\ie, numerically unstable) 
due to the sharp spectral functions (with small widths) in the parton propagators which compromises the numerical accuracy. Nevertheless, 
the sign of $ \chi^{uc}_{22}/\chi_2^c$ appears to turn negative at low temperature; $ \chi^{uc}_{11}/\chi_2^c$ is more stable and of the same sign 
as in the SCS plotted in Fig.~\ref{fig_sus}, albeit of smaller magnitude. Further scrutiny of this finding is in order to better establish 
the degree to which heavy-light susceptibilities are sensitive to the underlying color forces.

\section{Conclusions and Perspective}
\label{sec_con}
Employing a thermodynamic $T$-matrix approach to the QGP extended to finite chemical potential, we have performed calculations of various 
quark-number susceptibilities in a  strongly-coupled scenario. Toward this end we have introduced two additional model parameters into our 
framework which quantify the leading order corrections in $\mu_q$ to the screening mass of the interaction kernel and the bare light-parton masses 
of the bulk medium. They have been fit to reproduce the temperature dependence of the lQCD data for the second-order baryon susceptibility, 
$\chi_2^B$. The additional screening of the in-medium potential at finite $\mu_q$ turns out to be rather moderate. The resulting diagonal charm-quark susceptibilities,  $\chi_2^c$ and $ \chi_4^c $, which do not depend on the additional fit parameters and thus can be considered  predictions of our 
approach, show fair agreement with the lQCD results. In particular, the large collisional widths inherent in the charm-quark spectral functions, as 
well as bound-state correlations close to $T_{\rm pc}$, are instrumental in this agreement (as demonstrated by a calculation with quasiparticle 
charm quarks, which falls short of the lQCD data).      
We have also computed the off-diagonal charm susceptibilities, $ \chi^{uc}_{11}/\chi_2^c$ and $ \chi^{uc}_{22}/\chi_2^c$, for which very few results 
exist in the literature. Our calculations lead to fair agreement with pertinent lQCD data; most notably, when approaching $T_{\rm pc}$ from above, we find increasingly negative values of $ \chi^{uc}_{11}/\chi_2^c$, which to our knowledge has not been reported before in a microscopic model approach.
Our findings imply that a strongly-coupled system where large collisional widths are driven by the emergence of near-threshold resonances in attractive 
heavy-light color channels and produce a small HQ diffusion coefficient, $2\pi T D_s$$\simeq$2-5~\cite{Liu:2016ysz}, remains a viable realization of the 
sQGP at moderate temperatures.

Several future developments are in order to further scrutinize our understanding of these mechanisms. The current $T$-matrix formalism only accounts
for mesonic and diquark channels; while the latter is a building block of baryons,  the inclusion of genuine 3-body interactions remains to be elaborated, 
which is particularly interesting in view of charm-baryon production in nuclear collisions at the LHC and its implementation into recombination 
models~\cite{He:2019vgs}. 
Furthermore, the effects of spin-spin and spin-orbit interactions should be studied, which are dictated by a quantitative hadron spectroscopy in vacuum 
and are presumably essential to construct a smooth cross-over from partonic to hadronic bulk matter. Finally, the development in the present paper 
paves the way for deploying the $T$-matrix formalism into the finite-$\mu_q$ plane of the QCD phase diagram, where it could help to understand the 
microscopic interactions underlying the transport and spectral properties of QCD matter as produced in heavy-ion collisions at lower energies and in 
neutron stars and their mergers. Work in some of these directions is in progress. 
\\

\acknowledgments
This work was supported by the Strategic Priority Research Program of Chinese Academy of Sciences, Grant No. XDB34000000 (SYFL) 
and by the U.S.~National Science Foundation (NSF) through grant no.~PHY-1913286 (RR).

\bibliography{refcnew}

\end{document}